\documentclass[useAMS,usenatbib,usegraphicx]{mn2e}
\usepackage{times}

\newcommand{\simlt}{\lower.5ex\hbox{$\; \buildrel < \over \sim \;$}}
\newcommand{\simgt}{\lower.5ex\hbox{$\; \buildrel > \over \sim \;$}}
\newcommand{\be}{\begin{equation}}
\newcommand{\ba}{\begin{eqnarray}}
\newcommand{\ee}{\end{equation}}
\newcommand{\ea}{\end{eqnarray}}

\title[Spectral dating of high-$z$ stellar populations]
{Spectral dating of high-redshift stellar populations}
\author[I.~Ferreras \& S.~K.~Yi]
{Ignacio Ferreras$^{1,2}$\thanks{E-mail:ferreras@phys.ethz.ch} and 
  Sukyoung K. Yi$^1$\\
$^1$ Physics Dept. Denys Wilkinson Building, Keble Road, Oxford OX1 3RH\\
$^2$ Institut f\"ur Astronomie, ETH H\"onggerberg HPF D8, CH8093 Z\"urich, Switzerland
}
\begin{document}

\date{MNRAS Accepted}

\pagerange{\pageref{firstpage}--\pageref{lastpage}} \pubyear{2004}

\maketitle

\label{firstpage}

\begin{abstract}
Age derivation techniques for unresolved stellar populations at high 
redshifts are explored using the NUV spectrum of LBDS~53W091 ($z=1.55$) 
and LBDS~53W069 ($z=1.43$). The photometry and morphology of these
galaxies -- which are weak radio sources -- suggest they are early-type
systems, a feature that makes them ideal test beds for the analysis
of their ages and metallicities with the use of population synthesis models.
In the analysis that is based on {\sl simple} stellar population models,
we find a significant degeneracy between the derived 
ages and metallicities both in optical+NIR photometric 
and NUV spectroscopic analyses. This degeneracy is not so 
strong for LBDS~53W069. However even in this case the stellar
age cannot be constrained better than to a range roughly
encompassing one third of the age of the Universe at $z=1.43$  
(90\% confidence level).
We have explored several independent population 
synthesis models and consistently found similar results. 
Broadband photometry straddling the rest-frame 4000\AA\ break is also 
subject to a strong age-metallicity degeneracy. 
The use of realistic chemical enrichment assumptions
significantly helps in disentangling the degeneracy.
Based on this method, we derive the
average stellar age for both galaxies around 
$\langle t_\star\rangle\sim 3.6-3.8$~Gyr with better 
constraints on the youngest possible ages ($\sim 3$~Gyr at the
90\% confidence level).
The comparison with simple stellar population 
models suggest sub-solar metallicities ($\log Z/Z_\odot =-0.2$).
A composite model using chemical enrichment gives slightly higher  
metallicities in both galaxies ($\log Z/Z_\odot =-0.1$).
Given that the stellar component in galaxies forms over times
which are larger than a typical chemical enrichment timescale, we 
conclude that composite stellar populations must be used in all
photo-spectroscopic analyses of galaxies.
From the observational point of view, the most efficient (and feasible) 
way to set limits
on unresolved stellar populations comprises a combination of 
Balmer absorption lines along with either low SNR
rest frame NUV spectroscopy or accurate optical and NIR photometry.
\end{abstract}


\section{Introduction}

Estimating the age of the unresolved stellar populations observed
in galaxies represents a major challenge in our understanding of
galaxy formation. An ideal observation should allow us to infer
the star formation history of galaxies from a set of various
spectrophotometric observables. The light from a recent burst is
dominated by the colour of OB stars, whereas the light from
old stellar populations come predominantly from G- and K-type
giants. Hence, as a zeroth order approximation, broadband colours
track the stellar age. However, this stellar clock is not good
enough because of the effect of metallicity on age estimates so
that -- within the observational uncertainties -- the colours of
galaxies, whose light is dominated by old and metal-poor stars, may
be indistinguishable from galaxies mainly composed of young and
metal-rich stars. Furthermore, this degeneracy cannot be simply
overcome by a better spectral resolution as targeted spectral indices
such as the Lick/IDS system (Worthey et al. 1994) suffer from an
age-metallicity degeneracy similar to broadband photometry (Worthey 1994).

Accurate spectroscopic dating of stellar populations has been
attempted over the past years (Stockton, Kellogg \& Ridgway 1995;
Dunlop et al. 1996; Spinrad et al. 1997; Yi et al. 2000) although 
a comprehensive analysis of the effect of
metallicity has not been considered in detail until
recently (Nolan et al. 2003; Ferreras, in preparation).

This paper focuses on LBDS~53W069 ($z=1.43$)
and LBDS~53W091 ($z=1.55$), two faint radio galaxies
from the Leiden-Berkeley Deep Survey (hereafter LBDS;
Windhorst et al. 1984a; 1984b).
The search for optical counterparts to
faint radio emission is a technique which should allow us to
spot old stellar populations at high redshifts (Kron, Koo
\& Windhorst 1985). However, the light from many of these objects
is dominated not by starlight but by the active nucleus.
An attempt was made to target old stellar populations
by searching in the LBDS catalog for weak radio sources
with faint NIR magnitudes ($K\leq 18$) and red optical$-$IR colors
($R-K>5$). Among the reddest galaxies, LBDS~53W069 ($K=18.5$; $R-K=6.3$)
and LBDS~53W091 ($K=18.7$; $R-K=5.8$) have been extensively studied 
(Dunlop et al. 1996; Spinrad et al. 1997; Dunlop 1999).
Spinrad et al. (1997) presented
the spectrum of 53W091 using Keck LRIS, which maps into its rest-frame NUV
($1960<\lambda/$\AA\ $<3500$). Based on a comparison with their 
simple stellar populations (SSPs), they found a minimum age
of $3.5$~Gyr, which imposed a significant constraint on cosmology.
However, due to the substantial difference between SSP 
models (e.g., Yi et al. 2003) and the systematic effects in age
derivation techniques, the subsequent analyses of 
Bruzual \& Magris (1997), Heap et al. (1998) and of Yi et al. (2000) 
indicated significantly younger ages ($\sim 1-2$~Gyr).
The controversy has continued. 
Nolan et al. (2003) have recently performed a more detailed analysis
exploring composite stellar populations (i.e. a mixture of metallicities)
and found $\sim 3$~Gyr roughly confirming their first age
estimate. The discrepancy between these two age estimates can
be translated into a formation redshift.
By adopting a $\Lambda$CDM 
cosmology with $\Omega_m=0.27$ and $H_0=71$~km~s$^{-1}$~Mpc$^{-1}$
(Spergel et al. 2003), which is used in this paper hereafter, the literature 
suggests that the stellar component of LBDS~53W091
could have been formed at a redshift between $z_F\sim 2.5$ (Yi et al. 2000)
and $z_F\simgt 4.7$ (Nolan et al. 2003).

The controversy over the actual age of an allegedly simple case
such as LBDS~53W091 shows that it is imperative to make a robust
estimate of the underlying uncertainties. 
We will explore in this paper the uncertainties inherent to any
age estimate using simple fitting techniques to the observed
NUV spectra of LBDS~53W069 and 53W091. Given the importance 
of these estimates
to cosmology as well as to galaxy formation, we believe the case
study presented in this paper is a timely and relevant exercise
which should be borne in mind when extracting ages from the
integrated properties of stellar populations.

\begin{figure}
\includegraphics[width=3.4in]{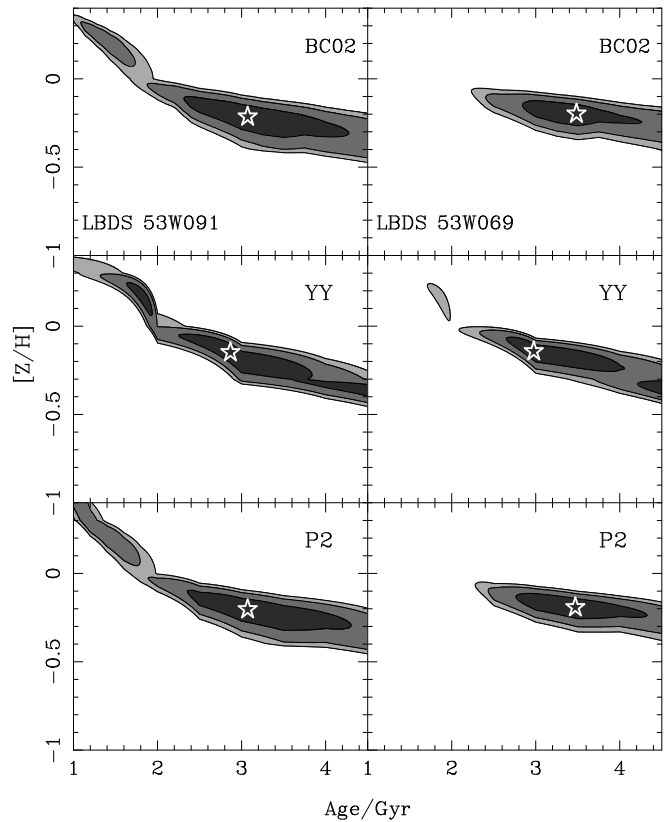}
\caption{$\chi^2$ map for a comparison of the observed SED of
galaxies LBDS~53W091 ($z=1.55$; left) and LBDS~53W069 ($z=1.43$) 
with simple stellar populations from three independent population
synthesis models (see text for details). In each panel
the grey scales correspond to the $1$, $2$ and $3\sigma$
confidence levels from centre to outside. The stars give the
position of the minimum $\chi^2$ (see Table~\ref{tab:ages}).}
\label{fig:c2sed}
\end{figure}

\section{Stellar dating using SSPs}

\begin{figure}
\includegraphics[width=3.4in]{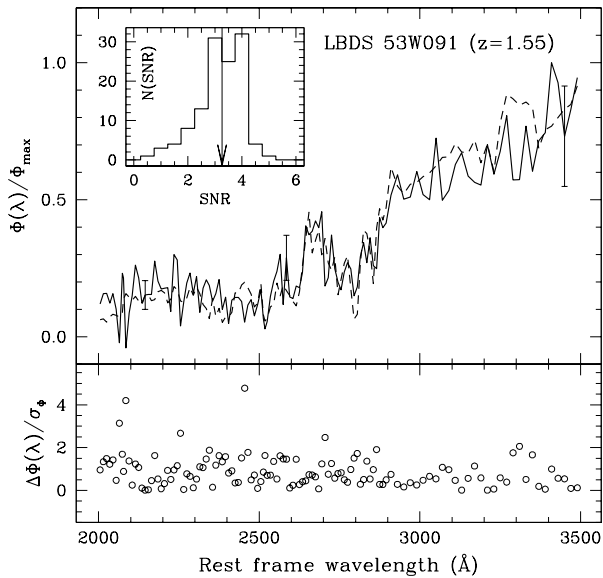}
\caption{SED of LBDS~53W091 ($z=1.55$) show in rest-frame wavelengths (solid line).
Three characteristic ($1\sigma$) error bars are shown. The dashed line is the
synthetic SED from the simple stellar population
which corresponds to the best fit from the YY models
(see \S2 for details). The inset shows the histogram of the observed signal-to-noise
ratios. The arrow give the position of the median. The bottom panel shows the
residuals of the fit as a fraction of the noise level. The reduced $\chi^2$
that corresponds to this fit is shown in Table~\ref{tab:ages}.
}
\label{fig:091}
\end{figure}

\begin{figure}
\includegraphics[width=3.4in]{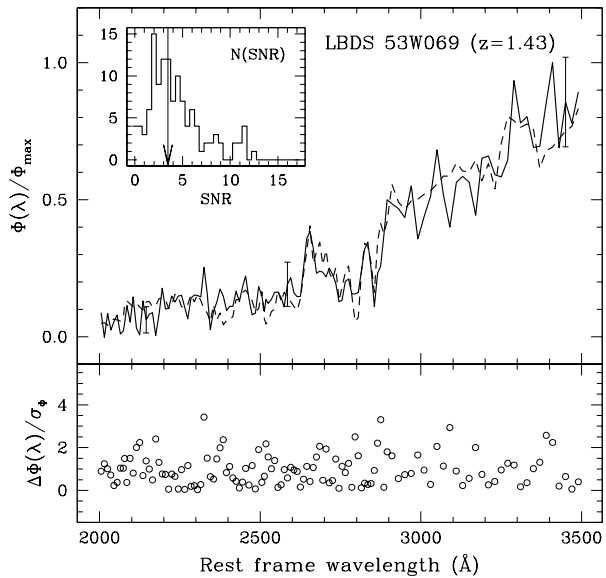}
\caption{Same as figure~\ref{fig:091} for galaxy LBDS~53W069 ($z=1.43$).}
\label{fig:069}
\end{figure}

A first approach to this problem involves the assumption that all
stars in a given galaxy have the same age and metallicity. These
so-called simple stellar populations (SSPs) constitute the
building blocks of population synthesis models, and they give good
fits to globular cluster data (e.g., Bressan, Chiosi, \& Fagotto
1994; Yi et al. 2001). Both galaxies targeted in this paper appear
morphologically to be early-type galaxies with a de~Vaucouleurs
profile in rest-frame $B$ and $R$ bands (Waddington et al. 2002).
Their optical colours hint at the presence of moderately old stars
(Spinrad et al. 1997). We have performed a first test comparing the
observed NUV spectrum with a grid of SSPs over a large range of
ages ($1<$\,t/Gyr\,$<4.5$) and metallicities 
($-1<$\,$\log Z/Z_\odot$\,$<+0.4$). The
age of the Universe at the redshift of these galaxies ($z\sim
1.5$) is 4.35~Gyr with the cosmology adopted in this paper.
Figure~\ref{fig:c2sed} shows the likelihood contours when
performing a $\chi^2$ test, taking into account the observed flux
and signal to noise ratio in the rest-frame range
$2000<\lambda/$\AA\ $<3500$. The three shaded areas represent 
(from dark to light grey) the $1$, $2$ and $3\sigma$ confidence levels.
When computing the $\chi^2$ measurements, we normalised both 
the observed and model spectra by their integrated flux in the 
wavelength range specified above. Each
panel corresponds to a comparison with different population
synthesis models. From top to bottom: BC02 for the latest 
``pre-STELIB'' Bruzual \& Charlot (1993) models; 
YY for Yi and Yoon (in preparation) which is an
updated version of the Yi, Demarque, \& Oemler (1997) models, and 
P2 for P\'egase~2 (Fioc \& Rocca-Volmerange 1997).
The stars give the position of the best fit for 
LBDS~53W091 ({\sl left}) and 53W069 ({\sl right}), also shown in
Table~\ref{tab:ages}, along with the best value of the reduced
$\chi^2$. The table gives the marginalized error bars for the age
and metallicity estimates at the 90\% confidence level. 
The final number of spectral data points 
in the analysis is $122$ (25\AA\, resolution) in both galaxies. 
Figures~\ref{fig:091}
and \ref{fig:069} show the observed SEDs of LBDS~53W091 (Spinrad et al. 1997) 
and LBDS~53W069 (Dey et al. in preparation), respectively. 
Three characteristic ($1\sigma$)
error bars are shown for reference. The inset gives the histogram of
signal-to-noise ratios, with the median shown by an arrow.The dashed 
line is the synthetic SED from the SSP which corresponds to the best 
fit from the YY models. The bottom panel shows the
residuals of the fit as a fraction of the noise level.

One can see that all models give similar results for both galaxies
NUV spectral fittings. 
The figure shows that the confidence levels are too wide
to pin down accurate age estimates; and, thus, the wide range
of ``best'' ages and metallicities shown in Table~\ref{tab:ages} is
to be expected. Hence, the most important conclusion one can
extract from Figure~\ref{fig:c2sed} is that the NUV spectrum used
for the test cannot give an accurate age estimate in a simple
comparison with SSPs because of 
the age-metallicity degeneracy. 
This degeneracy is smaller for LBDS~53W069, giving a reasonable
constraint on the stellar ages. However, notice that at the
90\% confidence level, one cannot rule out ages which
span roughly one third of the age of the Universe at the observed
redshift. 


\begin{table}
\caption{Best ages and metallicities (90\% confidence level)}
\label{tab:ages}
\begin{center}
\begin{tabular}{c|c|ccc}
Galaxy & Model & Age/Gyr & $\log Z/Z_\odot$ & $\chi_{\rm r}^2$\\
\hline\hline
LBDS~53W091 & BC02   & $3.07^{1.21}_{-0.89}$ & $-0.21^{0.14}_{-0.14}$ & $1.39$\\
            & YY$^a$ & $2.87^{1.45}_{-1.11}$ & $-0.15^{0.34}_{-0.21}$ & $1.40$\\
            & P2     & $3.07^{1.19}_{-1.49}$ & $-0.20^{0.34}_{-0.14}$ & $1.39$\\
            & CSP$^b$& $3.62^{0.24}_{-0.53}$ & $-0.09^{0.21}_{-0.32}$ & $1.36$\\ 
\hline
LBDS~53W069 & BC02   & $3.47^{0.88}_{-0.65}$ & $-0.20^{0.07}_{-0.09}$ & $1.49$\\
            & YY     & $2.98^{1.43}_{-0.26}$ & $-0.14^{0.06}_{-0.20}$ & $1.52$\\
            & P2     & $3.47^{0.84}_{-0.68}$ & $-0.19^{0.07}_{-0.08}$ & $1.49$\\
            & CSP$^b$& $3.84^{0.31}_{-0.82}$ & $-0.08^{0.29}_{-0.33}$ & $1.43$\\
\hline
\end{tabular}
\end{center}
$^a$ Yi et al. (2000) found roughly 2~Gyr as the best fitting age assuming
the solar abundance. Here, we used the same models; but we get a larger 
age because lower metallicities are allowed.\\
$^b$ Composite models assuming a consistent chemical enrichment history.
Mass-weighted ages and metallicities are shown. See \S4 for details.
\end{table}

\begin{figure}
\includegraphics[width=3.4in]{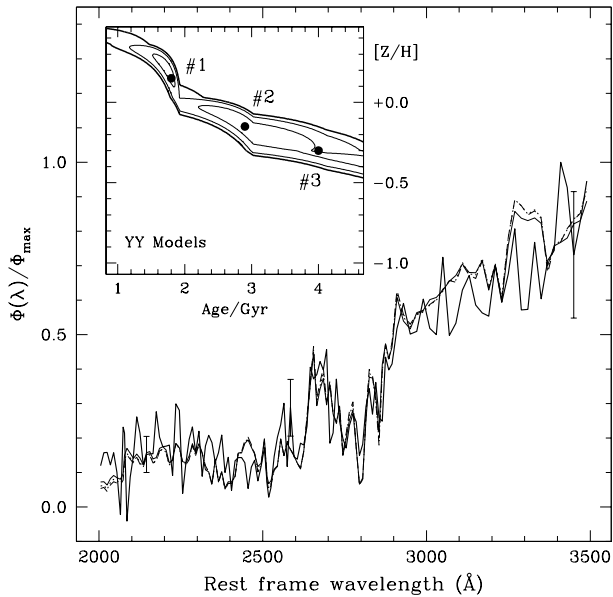}
\caption{A comparison of the observed Keck LRIS spectrum of LBDS
53W091 and three simple stellar populations from the $YY$ models
(Yi and Yoon in preparation). Three representative error bars are
shown. The inset shows contours of the $1$, $2$ and $3\sigma$ confidence
levels according to a $\chi^2$ test as discussed in the paper. Three points in
parameter space have been chosen to illustrate the wide range of
ages and metallicities which are compatible with the data. The SEDs
corresponding to points 1, 2 and 3 are shown as a thin solid, dotted
and dashed line, respectively.}
\label{fig:sed}
\end{figure}

Figure~\ref{fig:sed} further illustrates this point. The inset
shows a likelihood contour similar to those in
Figure~\ref{fig:c2sed} for the YY models. Three (age, metallicity)
pairs are chosen: \#2 corresponds to the best fit, \#1 and \#3 are
estimates along the ``likelihood ridge''.
The SEDs corresponding to the SSPs
for these three points are shown in the figure as thin solid, dotted and
dashed lines corresponding to points $1$, $2$ and $3$, respectively. The
observed SED of LBDS~53W091 is shown as a thick line with three representative
$1\sigma$ error~bars. One can see the SEDs corresponding to all three points
are nearly indistinguishable even though they span a large range of
both age and metallicity. 

\begin{figure}
\includegraphics[width=3.4in]{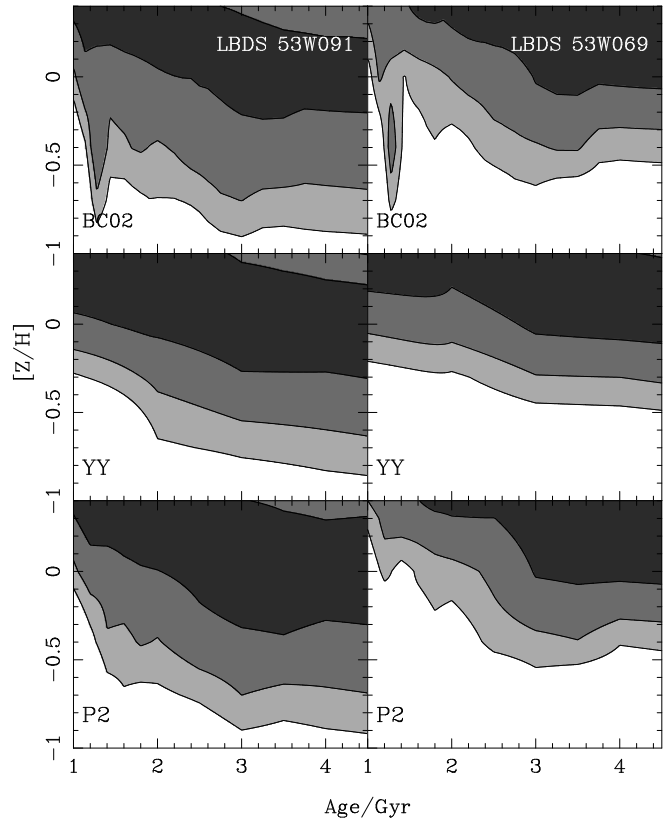}
\caption{Likelihood maps comparing the $i,J,H,K$ photometric data
of LBDS~53W091 ({\sl left}) and LBDS~53W069 ({\sl right}) from 
Waddington et al. (2000) with the predictions for simple
stellar populations from four independent modellers (see text 
for details). The grey shades correspond (from dark to light grey) 
to the $1$, $2$ and $3\sigma$ confidence levels.}
\label{fig:c2phot}
\end{figure}

We have also explored the constraints one could set on a
comparison with SSPs using the broadband photometric data
from Waddington et al. (2000). Figure~\ref{fig:c2phot} shows
likelihood contours in age and metallicity when 
performing a $\chi^2$ test with the $r-i$, $i-J$, $J-H$ and $H-K$
colours. 
We excluded the Gunn-$g$ data from the analysis because of its
large error bars. Through this exercise, we also demonstrate the 
difference between the analysis on the low-SNR SED and 
the one on the broadband photometry with higher SNR.
The shades (from dark to light grey)
correspond to the $1$, $2$, and $3\sigma$ confidence levels for galaxies
LBDS~53W091 ({\sl left}) and 53W069 ({\sl right}), respectively. 
One can see that the
age-metallicity degeneracy is stronger than in the case of a
comparison with the NUV SED. 

Dust was not explored in this comparison. In this sense, these age
estimates could be considered as upper limits, since the reddening
from dust will make a stellar population appear older when analyzed
by a dustless model. Yi et al. (2000) estimated that a modest amount
of reddening -- E($B-V$)=0.04~mag -- can reduce the age by about
$0.5$~Gyr, reconciling their spectroscopic and photometric dating
of LBDS~53W091.
Furthermore, a small error in the flux 
calibration of the SED will change the overall shape of the spectrum,
with a significant change in the age and metallicity obtained in this
way. We want to emphasize here that an accurate spectral dating requires
a very precise flux calibration.
We conclude in this section that a comparison of
the available data from LBDS~53W069 or 53W091 and SSPs over a wide
range of ages and metallicities cannot give us an accurate value
of the age unless an independent estimate of the metallicity is
obtained. In the next section we explore on possible ways to
improve the age dating of unresolved stellar populations at high
redshift.

\section{How to improve the age estimate}

\subsection*{I. Rest-frame NUV spectroscopy at higher SNR}
\begin{figure}
\includegraphics[width=3.4in]{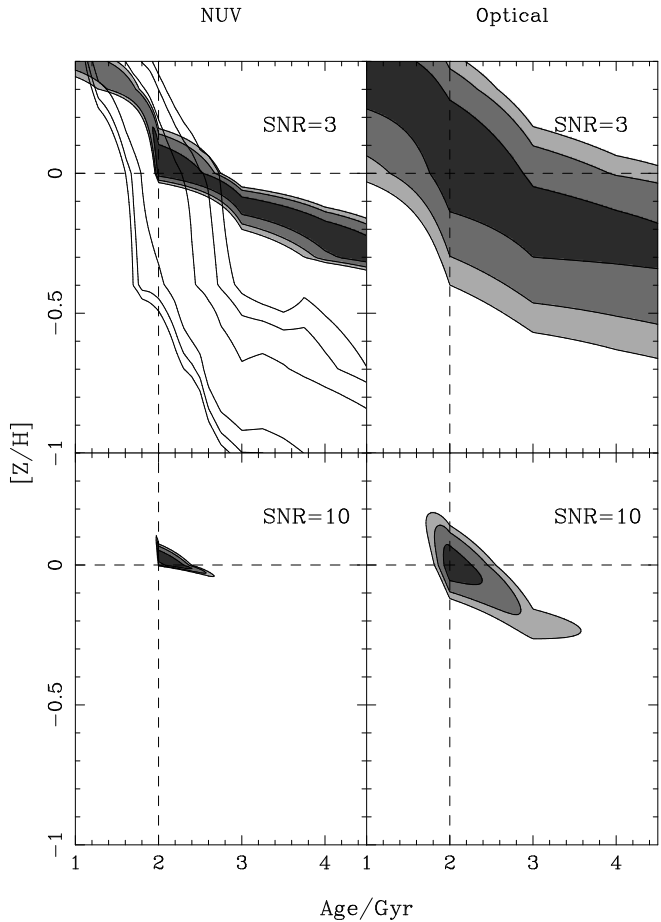}
\caption{Comparison of a synthetic SED from a SSP of the $YY$
models for $t=2$~Gyr and solar metallicity (represented by the
dashed lines). The $\chi^2$ test is performed using a grid of SSPs
from the BC03 population synthesis model. The contours are shown
(from dark to light grey) at the $1$, $2$, and $3\sigma$
confidence levels. The synthetic SEDs are generated at the
same redshift as LBDS~53W091 (i.e. $z=1.552$) and for two
different signal-to-noise ratios, namely SNR=3 ({\sl top}) --
which roughly corresponds to that of the observed spectrum of
LBDS~53W091 (see Figure~\ref{fig:091}) -- and SNR=10 ({\sl bottom}). 
We have analysed two
sets of SEDs corresponding to two different spectral regions:
near-ultraviolet (NUV, {\sl left}) with the same spectral range as
the observed one, and optical (OPT, {\sl right}). The top-left panel
also shows the $1$, $2$, and $3\sigma$ confidence levels of a 
putative measurement of Balmer absorption, namely $H\beta\sim 2.8\pm 0.1$\AA\  ,
(rest-frame), expected for the fiducial age ($2$~Gyr) and solar metallicity 
assumed for this synthetic SED. See text for details.
}
\label{fig:syn}
\end{figure}

Obtaining spectra of high-redshift galaxies requires very long
integration times on large telescopes. For instance, the 
rest-frame NUV SED of LBDS~53W091 shown in Figure~\ref{fig:091}
required an effective exposure time of $\sim
20,000$~s using LRIS at the 10~m W.~M.~Keck telescope (Spinrad et
al. 1997). The spectrum has a median SNR$\sim 3.5$ per 
resolution element (25\AA). We decided to
explore the effect of a higher signal to noise ratio on the
estimates of the age and metallicity. In the left panels of
Figure~\ref{fig:syn} we generated a {\em synthetic galaxy spectrum} with the
same spectral coverage and resolution as that of LBDS~53W091, at
SNR$=3$ ({\sl upper}) and SNR$=10$ ({\sl lower}). The likelihood
map was obtained for a set of 200~realizations of the spectrum
corresponding to a SSP from the YY models for a $t=2$~Gyr and
$Z=Z_\odot$ (these fiducial values are shown in the figure as a
dashed line). The shaded contours -- from dark to light gray --
correspond to the $1$, $2$ and $3\sigma$ confidence levels. 
Obviously, a more time-demanding spectrum at a higher SNR
will reduce the width of the ``likelihood ridge'' although it
could not break the age-metallicity degeneracy unless very high
SNRs ($\sim 10$) are achieved -- which would imply prohibitively 
long exposure times even for a 10~m-class telescope. 
Hence, for all practical purposes, 
the accuracy of the age would still hinge on an independent 
estimate of the metallicity.

\subsection*{II. Spectroscopy across 4000\AA\  break}
The strong age sensitivity of the 4000\AA\  break along with 
its weaker dependence with metallicity for young stellar
populations (see e.g. Kauffmann et al. 2003)
suggests that a spectral coverage encompassing this
break could be a good age indicator. 
Therefore, we decided to test if optical SEDs covering the 
4000\AA\ break lead to more refined age estimates than NUV SEDs.
The right panels
of Figure~\ref{fig:syn} explore this possibility. We generated
200~synthetic spectra corresponding to a SSP with $t=2$~Gyr and
$Z=Z_\odot$ from the YY models. These SEDs were computed for SNR$=3$
({\sl top}) and SNR$=10$ ({\sl bottom}). The spectral resolution
was assumed to be the same as that for the observed rest-frame NUV
SED (i.e. $\Delta\lambda =25$\AA\ ) and the spectral coverage was
chosen to straddle the 4000\AA\ break: $9000<\lambda /$\AA\
$<16500$ in the observer's frame, which corresponds to
$3525<\lambda /$\AA\  $<6450$ in the rest-frame. The figure shows
that one does not achieve a better age constraint at all by using 
the optical SED straddling the 4000\AA\ break, compared to 
using the rest-frame NUV SED. However, the SNR=10 
optical SED yields age estimates with markedly better accuracy 
compared to the SNR=3 rest-frame NUV SED.
For 1--5\,Gyr populations, rest-frame optical 
SEDs should take shorter exposures by an order of magnitude
to achieve the same SNR and but require higher values of SNR.
It is encouraging to find that high SNR rest-frame optical 
SEDs can be effective tools for deriving ages even at low 
spectral resolution (25\AA).

\subsection*{III. Broadband photometry across 4000\AA\  break}
One alternative way of estimating the ages from the flux across
the 4000\AA\  break would be to perform broadband photometry. This
is much less time-consuming than spectroscopy and it is questionable
whether one can do better than shown in Figure~\ref{fig:c2phot} if 
a more accurate photometry is performed on these galaxies.
Figure~\ref{fig:synphot} shows the
likelihood map using a synthetic SSP from the YY models for
t$=2$~Gyr and $Z=Z_\odot$ assuming a small photometric error in the
colours ($\pm 0.05^m$). We targeted $V-R$ and $R-K$
colours, which map into NUV and optical rest-frame
colours, respectively. The shaded regions are -- from dark to
light grey -- the $1$, $2$ and $3\sigma$ confidence levels. 
The two upper panels show the analysis when photometry is used
in the analysis: only $V-R$ colour ({\sl top}), or when both 
colours ($V-R$ and $R-K$) are considered ({\sl middle panel}). 
This approach fares equally well compared to the more time-consuming
continuum fitting at low SNRs. The age-metallicity degeneracy still
persists.

\subsection*{IV. Narrow spectral indices}
\begin{figure}
\includegraphics[width=3in]{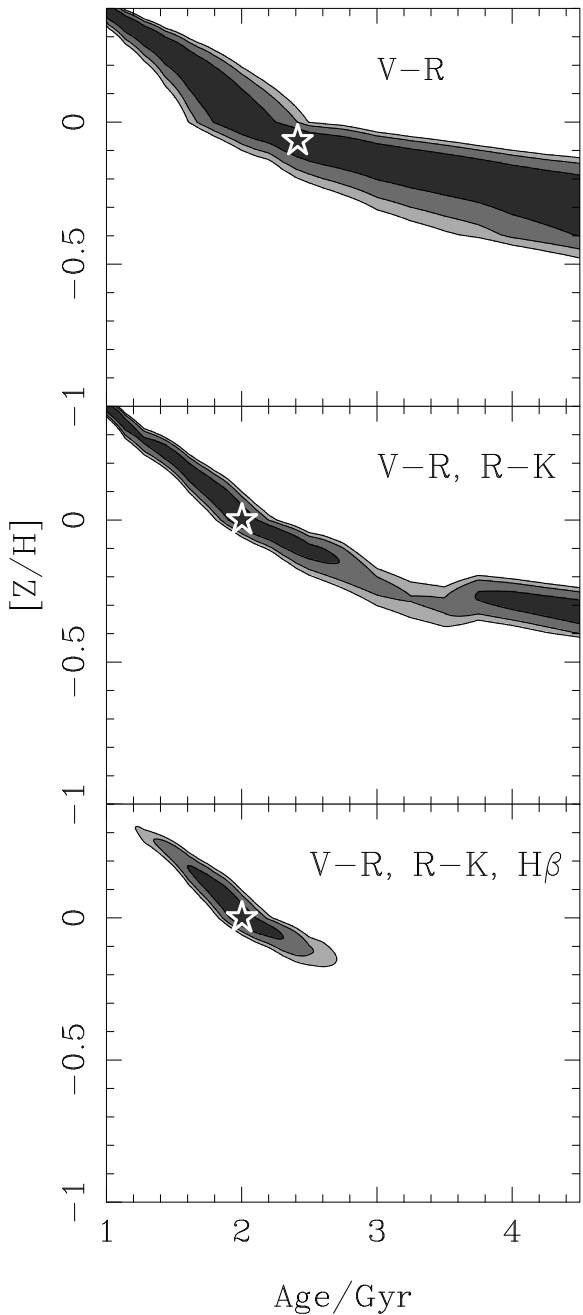}
\caption{The contour plots correspond to an estimate of age and metallicity
using broadband photometry plus a Balmer index. From top to bottom the
observational constraints are $V-R$ only; $V-R$ along with $R-K$ or 
these two colours plus H$\beta$ absorption. The photometric uncertainty
in this simulation is $\pm 0.05$~mag. H$\beta$ is assumed to be measured
with a $\pm 0.1$\AA uncertainty in the rest frame of the simulated galaxy
($z=1.55$). The stars show the position of the best fit. The simulated
galaxy has an age of $2$~Gyr and solar metallicity.}
\label{fig:synphot}
\end{figure}

After exploring the various approaches discussed above, we are left
with the option of targeting narrow spectral features. This is a
technique often used in age estimates of early-type galaxies (e.g.
Trager et al. 2000; Kuntschner 2000; Bernardi 2003). The key issue
is to target line indices which have a significantly different
dependence on age and metallicity. As a simple test, we have 
checked the likelihood maps one would get from a measurement of a
Balmer index such as H$\beta$. Balmer indices are very prominent
in stellar atmospheres at $T\sim 10,000$~K, which corresponds to
main sequence A-type stars. Hence, Balmer absorption is especially
strong in stellar populations over an age range $1-3$~Gyr, which
is ideal for $z\simgt 1.5$ galaxies. Old, metal-poor stars can
also contribute significantly to Balmer absorption lines (Lee,
Yoon, \& Lee 2000). However, high-redshift galaxies are immune to
the complication because they are still too young to have
developed such stars. 
The lines in the top-left panel of Figure~\ref{fig:syn} represent 
the $1$, $2$ and $3\sigma$ confidence levels of a simulated measurement
of Balmer absorption corresponding to H$\beta =2.8\pm 0.1$\AA\   in the
rest frame. This is the value to be expected for the fiducial model
targeted in the analysis throughout \S3 (i.e. an age of 2~Gyr and solar 
metallicity). One can see that adding Balmer absorption to the analysis 
helps in constraining the ages. 
The contour levels in the bottom panel of 
Figure~\ref{fig:synphot} show similar confidence levels
for a combined measurement of broadband photometry (using colours
$V-R$ and $R-K$) along with the same value of H$\beta$ as above. 
We assume a higher photometric accuracy $\pm 0.05$~mag than 
currently available, i.e. $\pm 0.1-0.2$~mag (Waddington et al. 2000).
Notice that the use of accurate photometry imposes 
similar constraints on the age and metallicity as in the combined
analysis of the NUV SED at low signal-to-noise ratios shown
in the Figure 6 {\sl top-left} panel.
Hence, a combination of broadband photometry with 
moderate resolution spectroscopy targeting age-sensitive 
indices such as H$\beta$ is the best observational approach to 
an accurate estimate of the age of LBDS~53W091.

We have chosen H$\beta$ to illustrate the usefulness of Balmer
line measurements, but higher order Balmer line indices may be 
preferred especially depending on the redshift of the target galaxy. 
The recent study of van Dokkum and Ellis (2003) presents a clear 
example of such successes.

\section{The effect of chemical enrichment}
So far, we have based our analysis of stellar ages and metallicities
on simple stellar populations. Age estimates based on SSPs rely on the
assumption that the stars have the same age and metallicity.
Globular clusters are the best candidates for a realistic
SSP, as inferred by numerous colour-magnitude diagram analyses. 
Various spectrophotometric properties of
early-type galaxies hint at a very fast star formation process
implying a rather small age range (e.g. Bower, Lucey \& Ellis 1992;
Kuntschner 2000). However, the radial colour gradients found
in elliptical galaxies (e.g. Peletier et al. 1990) 
implies that the stellar component is
distributed over a significant range of metallicities.
Furthermore, low-mass ellipticals both nearby and
at moderate redshifts display rest-frame NUV colours (Ferreras \& Silk 2000a),
Balmer absorption, and [Mg/Fe] abundance ratios (Trager et al. 2000), all of
which are indicative of a more extended star formation history
(Ferreras \& Silk 2003). Hence, comparing early-type galaxies with SSPs
may not be such a good approximation.

Furthermore, the morphological analysis of LBDS~53W091 performed on HST
images using WFPC2 and NICMOS in the F814W, F110W and F160W passbands
is suggestive of a two component system, comprising a compact
de~Vaucouleurs spheroid ($r_e=2.7\pm 0.7$~kpc) plus an extended
exponential disk ($h=4.2\pm 1.7$~kpc),
which only appears in the bluer passband (Waddington et al. 2002).
This is suggestive of the presence of young stellar populations.
The contamination in the rest-frame NUV from young stars
is estimated to be up to 20\% (Waddington et al. 2002).

Therefore, we perform a similar analysis as the one described in \S 2
using a chemical enrichment model which results in a (consistent)
distribution of ages and metallicities. Let us briefly describe
the parameters which determine the star formation history:
\begin{itemize}
\item[$\bullet$] The star formation rate ($\psi$) is determined by a power law, with
  a star formation efficiency ($C_{\rm EFF}$) which is fixed in this analysis.
  For simplicity, we decided to use a linear star formation law, namely:
  $\psi (t)=C_{\rm EFF}\rho_g(t)$, where $\rho_g$ is the gas mass.
\item[$\bullet$] Infall of pre-enriched gas is assumed, at a metallicity
  $Z_\odot /10$. The infall rate follows a generic ``delayed
  exponential'' profile: $f(t) \propto \Delta t \exp (-\Delta t^2/2\tau_f^2)$,
  where $\Delta t = t - t(z_F)$, with $z_F$ being a ``formation redshift'',
  and $\tau_f$ is the infall timescale.
\item[$\bullet$] Gas outflows are parametrized by $0\leq B_{\rm OUT}\leq 1$, which
  defines the fraction of the gas returned from stars which is ejected 
  from the galaxy.
\end{itemize}
The model tracks the stellar, gas and metal components in a single zone.
The rest of the details follow standard assumptions about
stellar evolution and chemical enrichment. We assume a Salpeter (1955)
initial mass function in the $0.1<M/M_\odot <60$ mass range.
The chemical enrichment model is described in detail
elsewhere (see e.g. Ferreras \& Silk 2000b).
With this generic parametrization we ran for each galaxy 
a set of $32\times 32\times 32$ star formation histories encompassing a 
wide range of outflow fractions ($0\leq B_{\rm OUT}\leq 1$); 
formation times ($0.4\leq t(z_F)/{\rm Gyr}\leq 4.2$), and 
infall timescales ($0.05\leq\tau_f/{\rm Gyr}\leq 2$). The upper limit
chosen for $\tau_f$ is motivated by the fact that longer infall
timescales will result in a significant fraction of ongoing star 
formation at the observed redshift. Hence, models with 
$\tau_f\simgt 2$~Gyr have
a strong NUV component from very young stars which is incompatible
with the observed SEDs of our galaxies. The star formation efficiency was
fixed at a high value ($C_{\rm EFF}=50$ Gyr$^{-1}$) after checking that
low star formation efficiencies were consistently giving higher values
for $\chi^2$. This value of the efficiency is consistent with similar 
models of star formation in elliptical galaxies (Ferreras \& Silk 2000b).

Yi et al. (2000) and 
Nolan et al. (2003) already considered a simple analysis of 
stellar populations with mixed metallicities, although their models did not
assume a consistent age-metallicity relation obtained by a proper treatment
of chemical enrichment. 
Yi et al. (2000) adopted metallicity distributions from instant starburst
models (and thus no age spread). 
Nolan et al. (2003) on the other hand randomly combined populations
from seven metallicity bins from $0.01$ to $5Z_\odot$. In our models,
we consistently evolve the metallicity according to the prescriptions
described above. 
It is important to note that our star formation history (age-metallicity
relations) is consistent with the colour-magnitude relations (CMR)
of early-type galaxies at the present epoch (Bower et al. 1992).
In other words, our models are not randomly selected but instead
calibrated to the local CMR information.

\begin{figure}
\includegraphics[width=3in]{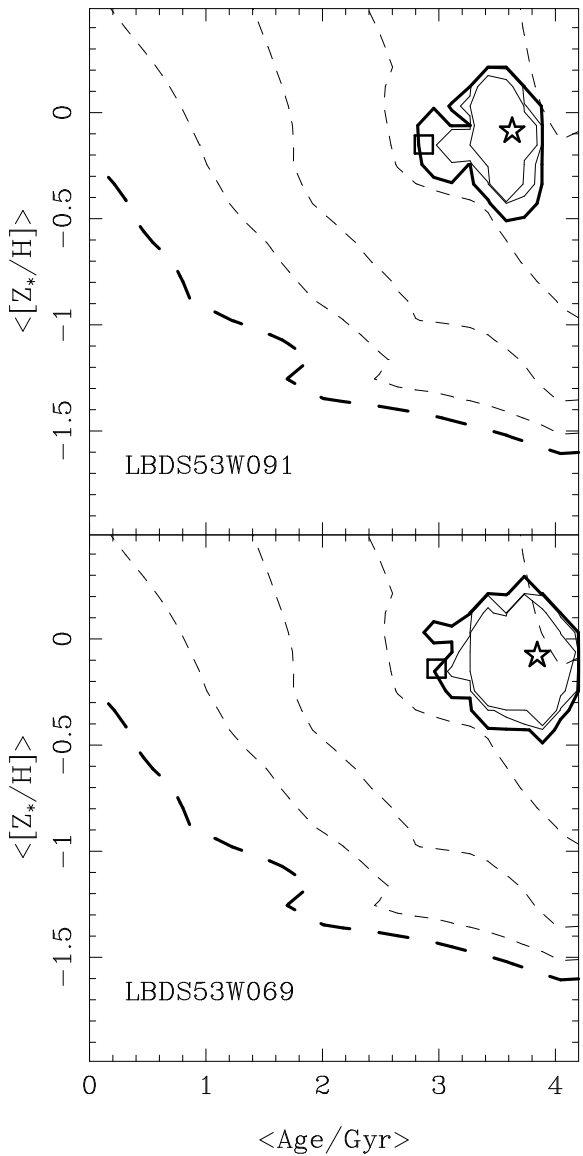}
\caption{Likelihood contours comparing the observed NUV SED of
LBDS~53W091 ({\sl top}) and LBDS~53W069 ({\sl bottom}) with a set
of composite models following a chemical enrichment model (see text for
details). The contours are at the $1$, $2$ and $3\sigma$ (thick line) 
confidence levels as a function of the (mass-weighted) age and 
metallicity of the stars. The star gives the position of the 
best fit and the square is the best fit for a comparison with
simple stellar populations (see \S2).
The dashed lines are contours of the
average value of the Balmer H$\beta$ index from 2\AA\ ,
in the upper right corner to 4\AA\  (thick line), in steps 
of 0.5\AA\ . The model predictions fall in the range of
$2\leq $H$\beta$/\AA\ $\leq 2.5$.
}
\label{fig:c2chem}
\end{figure}

We have performed a $\chi^2$ test on these composite models compared
to the observed (SNR$\sim 3$) NUV SEDs of the two galaxies.
Figure~\ref{fig:c2chem} shows the marginalized likelihood contours
as a function of the average (mass-weighted) age and metallicity. The
solid lines give the $1$, $2$ and $3\sigma$ (thick) confidence levels.
The best fits, uncertainties and reduced chi-squares for these composite
models are also given in table~\ref{tab:ages} under CSP.
One can see that the use of a chemically consistent model helps in 
constraining the mean metallicity effectively so that the age estimates
can be better determined. The $3\sigma$ confidence level 
limits the average
age between $2.8$ and $4$~Gyr, with an average (mass-weighted) metallicity
around solar, with an uncertainty around $\pm 0.5$~dex.
Our result is in agreement with the randomly
mixed metallicity model used in Nolan et al. (2003), whose models give 
a stellar age around $3$~Gyr.

It is important to note that we achieve larger age estimates from the
NUV spectral fitting analysis when we use $composite$ models than 
when we use $SSPs$. The star and square in figure~\ref{fig:c2chem}
give the position of the best fit for the analysis using composite
and simple stellar populations, respectively. The SSP analysis lies
at the $3\sigma$ limit of a more consistent approach using a
mixture of ages and metallicities.
Notice that the best fit for a composite model
corresponds to an average age around 3.6~Gyr with 0.8Z$_\odot$
(table~\ref{tab:ages}), whereas the SSP-based analysis shown in 
figure~\ref{fig:c2sed} would require $\sim 2$~Gyr if the assume the
same metallicity.
This result is a consequence of the mixture
of stellar populations with difference ages and metallicities, and
illustrates the fact that composite stellar population should be
used for these analyses. Furthermore, the best chemical enrichment model
gives an average age and metallicity which would be readily ruled
out in a naive approach using SSPs (see figure~\ref{fig:c2sed}).

One effective way to achieve a more accurate age estimate lies again in
the use of age-sensitive observables such as Balmer absorption.
The dashed lines in Figure~\ref{fig:c2chem} give contours of the 
{\sl hypothetical} H$\beta$ index measurements
from $2$\AA\  -- in the upper right corner -- to $4$\AA\  (thick) in 
steps of $0.5$\AA\ . The chemical enrichment model predictions 
correspond to values of H$\beta$ between $2$ and $2.5$\AA\ . 
Hence, Balmer absorption can significantly
reduce the uncertainties in the age estimate and check the validity of
the chemical enrichment models. Furthermore, we must emphasize here that 
all age estimates based on spectral fittings over a wide range of
wavelengths are heavily dependent on a precise calibration of the SED.
The overall shape of the SED plays a crucial role in all model
predictions presented in this paper. Therefore, it is very important to 
understand the uncertainties in the flux calibration of the spectra
to avoid large systematic errors in the analysis. Hence, the 
analysis of Balmer absorption should be considered as a valuable
cross-check in order to give an accurate answer for the stellar ages.

\section{Conclusions}

The observed rest-frame NUV SEDs of high redshift weak radio galaxies 
have been claimed to be robust estimators for the ages of old stellar
populations at high redshifts, which in turn allow us to set
constraints on the age of the Universe and on cosmological parameters
(Spinrad et al. 1997). However, in this paper
we show that the combined effect
of age and metallicity results in large error bars which are shown to
be independent of the population synthesis model used. 
This problem persists even if we use the spectral 
energy distribution over a wide range of wavelengths
instead of a set of broadband filters. Only at SNR=10 or greater
can the data 
disentangle the degeneracy. Unfortunately, this corresponds to
prohibitively long exposure times on a 10~m-class telescope.

A comparison of synthetic SEDs built from SSPs, with noise mimicking 
that of the observed data, shows that signal to noise ratios close 
to those used in the analysis of LBDS~53W091 (i.e. SNR$\sim$ 3) 
are not high enough to yield age estimates with appreciable precision. 
Slightly higher SNR will help in reducing the error bar, 
but the age-metallicity degeneracy is 
still very strong unless one can achieve a SNR=10 in rest frame
NUV spectroscopy or even higher in the optical spectral window. 
In principle, it may appear that the rest-frame NUV would be a 
desirable window to explore for these galaxies since it could
pose stronger constraints on
the allowed region of parameter space compared to a 
rest-frame optical SED. However, the weaker stellar continuum and the
strong effect of dust in this spectral region along with our
poorer knowledge of stellar emission in the NUV imply that it may
be more feasible and useful to obtain high precision rest-frame optical 
photometry straddling the 4000\AA\  break. Old stellar
populations are much brighter redward of the break, which results
in higher SNRs. 

Dating unresolved stellar populations from their spectral energy distribution
over a wide range of wavelengths is strongly dependent on the overall
shape of the continuum. A small error in the flux calibration will
distort this shape, thus altering the estimated ages and metallicities.
Hence, we want to emphasize that this type of studies requires SEDs with 
a very accurate flux calibration.
As has been widely known for a decade, the use of Balmer absorption 
lines, in combination with broadband photometry, largely solves 
the problem, and helps us in understanding a possible systematic effect
derived from uncertainties in the flux calibration of the SED.
As we discussed in \S 3, different concerns would
lead observers to choose different Balmer indices. At $z=1.5$ indices
such as H$\beta$, H$\gamma$ and H$\delta$ all appear in the NIR spectral
window, complicating an accurate ground-based spectroscopic
measurement. 

Hoping to break the infamous age-metallicity degeneracy from a 
theoretical point of view, we have explored a large set of
chemically consistent $composite$ population models.
Simply because no arbitrary metallicity is allowed in such
a scheme, this approach helped us determine metallicities
much better. We found a better constraint on the 
age estimates, giving a range of ages between $2.8$ 
and $4$~Gyr at the $3\sigma$ confidence level for both galaxies,
with a metallicity around solar with a $\pm 0.5$~dex uncertainty.
LBDS~53W069 seems to accept models with higher average metallicities.
Our results -- involving different sets of population synthesis models
and a detailed chemical enrichment scenario -- give similar 
results as the analysis of Nolan et al. (2003).
Even though any model of chemical enrichment introduces further
uncertainties in the modelling, galaxies should be considered
composite models as the star formation takes place over times which 
are always longer than the characteristic chemical enrichment
timescales. It is worth noticing that the best fits for a
simple and a composite stellar population are marginally compatible
(see figure~\ref{fig:c2chem}). Hence, at the expense of adding
further uncertainties to the modelling, we believe a proper
mixture of stellar ages and metallicities should be considered
in all photo-spectroscopic analyses of galaxies.


\section*{Acknowledgments}
We would like to thank James Dunlop and Louisa Nolan for very 
useful comments and for sending us the SEDs of the galaxies 
explored in this paper. We also thank Steve Rawlings, 
Pieter van Dokkum, Eric Gawiser and Hugues Mathis for 
useful discussions. This research has been supported 
by PPARC Theoretical Cosmology Rolling Grant PPA/G/O/2001/00016.

\bsp

\label{lastpage}

\end{document}